\documentclass[12pt,preprint]{aastex62}
\usepackage{epsfig}
\usepackage{graphicx}
\usepackage{color}
\usepackage{enumerate}
\usepackage{ulem}

\def\beq{\begin{equation}}
\def\eeq{\end{equation}}

\def\pc{{\rm\,pc}}

\def\yr{{\rm\,yr}}
\def\Myr{{\rm\,Myr}}

\def\kms{{\rm\,km\,s^{-1}}}
\def\kpc{{\rm\,kpc}}

\def\rbh{r_{\rm BH}}
\def\rt{r_{\rm t}}

\def\msun{\rm M_\odot}

\def\mstar{\rm m_*}
\def\rstar{\rm r_*}

\def\nbody{$\textit{N}$-body~}
\def\rinf{r_{\rm h}}
\def\Mbh{M_{\rm BH}}
\def\Mbhh{M_{\rm BH1}}
\def\Mbhl{M_{\rm BH2}}
\def\thetaD{\theta_{\rm D}}

\def\thetalc{\theta_{\rm{lc}}}
\def\rcrit{r_{\rm {crit}}}

\submitjournal{ApJ}
\shorttitle{TIDAL DISRUPTION RATE EVOLUTION FOR SMBHB}
\shortauthors{Li et al.}

\begin{document}

\title{DIRECT \textit {N}-BODY SIMULATIONS OF TIDAL DISRUPTION RATE EVOLUTION IN UNEQUAL-MASS GALAXY MERGERS}

\correspondingauthor{Shuo Li}
\email{lishuo@nao.cas.cn}

\author[0000-0001-6530-0424]{Shuo Li}
\affil{National Astronomical Observatories and Key Laboratory of Computational Astrophysics, \\
Chinese Academy of Sciences, 20A Datun Rd., Chaoyang District, Beijing 100012, China}
\affiliation{Department of Astronomy, School of Physics, Peking University, \\
Yiheyuan Lu 5, Haidian Qu, Beijing 100871, China}

\author{Peter Berczik}
\affil{National Astronomical Observatories and Key Laboratory of Computational Astrophysics, \\
Chinese Academy of Sciences, 20A Datun Rd., Chaoyang District, Beijing 100012, China}
\affiliation{Astronomisches Rechen-Institut, Zentrum f\"{u}r Astronomie, University of Heidelberg, \\
M\"{o}chhofstrasse 12-14, Heidelberg 69120, Germany}
\affiliation{Main Astronomical Observatory, National Academy of Sciences of Ukraine, \\
27 Akademika Zabolotnoho St., 03680 Kyiv, Ukraine}

\author{Xian Chen}
\affiliation{Department of Astronomy, School of Physics, Peking University, \\
Yiheyuan Lu 5, Haidian Qu, Beijing 100871, China}
\affiliation{Kavli Institute for Astronomy and Astrophysics, Peking University, \\
Yiheyuan Lu 5, Haidian Qu, Beijing 100871, China}

\author{F.K. Liu}
\affiliation{Department of Astronomy, School of Physics, Peking University, \\
Yiheyuan Lu 5, Haidian Qu, Beijing 100871, China}
\affiliation{Kavli Institute for Astronomy and Astrophysics, Peking University, \\
Yiheyuan Lu 5, Haidian Qu, Beijing 100871, China}

\author{Rainer Spurzem}
\altaffiliation{RS: Research Fellow at Frankfurt Institute for Advanced Studies}
\affil{National Astronomical Observatories and Key Laboratory of Computational Astrophysics, \\
Chinese Academy of Sciences, 20A Datun Rd., Chaoyang District, Beijing 100012, China}
\affiliation{Astronomisches Rechen-Institut, Zentrum f\"{u}r Astronomie, University of Heidelberg, \\
M\"{o}chhofstrasse 12-14, Heidelberg 69120, Germany}
\affiliation{Kavli Institute for Astronomy and Astrophysics, Peking University, \\
Yiheyuan Lu 5, Haidian Qu, Beijing 100871, China}

\author{Yu Qiu}
\affiliation{Department of Astronomy, School of Physics, Peking University, \\
Yiheyuan Lu 5, Haidian Qu, Beijing 100871, China}

\begin{abstract}

The hierarchical galaxy formation model predicts supermassive black hole binaries (SMBHBs) in galactic nuclei. Due to the gas poor environment and the limited spatial resolution in observations they may hide in the center of many a galaxy. However, a close encounter of a star with one of the supermassive black holes (SMBHs) may tidally disrupt it to produce a tidal disruption event (TDE) and temporarily light up the SMBH. In a previous work, we investigated with direct \nbody simulations the evolution of TDE rates of SMBHB systems in galaxy mergers of equal mass. In this work we extend to unequal mass mergers. Our results show that, when two SMBHs are far away from each other, the TDE rate of each host galaxy is similar as in an isolated galaxy. As the two galaxies and their SMBHs separation shrinks, the TDE rate is increasing gradually until it reaches a maximum shortly after the two SMBHs become bound. In this stage, the averaged TDE rate can be enhanced by several times to an order of magnitude relative to isolated single galaxies. Our simulations show that the dependence of the TDE accretion rate on the mass ratio in this stage can be well fitted by power law relations for both SMBHs. After the bound SMBHB forms, the TDE rate decreases with its further evolution. We also find that in minor mergers TDEs of the secondary SMBH during and after the bound binary formation are mainly contributed by stars from the other galaxy.

\end{abstract}

\keywords{Galaxies: evolution --- Galaxies: interactions --- Galaxies: kinematics and dynamics --- Galaxies: nuclei --- Methods: numerical}

\section{Introduction}
\label{intro}

Supermassive black hole binaries (SMBHBs) may form as a consequence of hierarchical galaxy mergers in the $\Lambda$ cold dark matter ($\Lambda$CDM) cosmology \citep{bege80,volo03}. Observations indicate that many massive galaxies harbor a supermassive black hole (SMBH) in the center, and there are close correlations between central SMBHs and their host galaxies \citep{mago98,ferr00,gebh00,trem02,korm13}. Because most of the massive galaxies may have experienced at least one major and many minor mergers in their evolution history, the existence of two or even multiple SMBHs in merger remnants should be expected. Taking into account the close link between SMBH and its host galaxy, the evolution of the SMBH pairs / SMBHBs in merging galaxies is of fundamental importance.

Whether two SMBHs in merging galaxies can sink down to the center of the remnant to form a bound binary, and how the binary evolves has been debated for many years. After a galaxy merger two SMBHs with their nuclear star clusters around them approach each other first through dynamical friction, until they form a gravitationally bound binary. But how the SMBHB evolves after that is not quite certain. If a SMBHB would shrink its orbit to several hundreds of Schwarzschild radii, it will finally coalesce due to energy loss by gravitational wave (GW) radiation in less than a Hubble time. Since dynamical friction is inefficient after the SMBHB became bound, additional processes such as few-body interactions between the SMBHB and surrounding stars or interaction with its gaseous environment are required to extract sufficient amounts of energy and angular momentum from the SMBHB to lead to the final coalescence \citep{bege80}. Otherwise the SMBHB may stall at a relatively large separation for more than the Hubble time. In a gaseous environment, a SMBHB could coalesce more easily\citep{goul00}, while in a gas poor environment, it has been argued that few-body superelastic scatterings between stars and the SMBHB are not efficient enough to lead to coalescence in less than a Hubble time \citep{bege80,milo03,ber05}. This has been named the "final parsec problem" in some publications; if true in general, SMBHBs should be ubiquitous in most galaxies. But it has been shown that the so-called "final parsec problem" is more like an artefact of special conditions in the model, such as strict spherical symmetry, no rotation, no tidal perturbations, no phases of gas inflow. However, both semi-analytical calculations and direct \nbody simulations later indicated that under realistic conditions (e.g. rotating or triaxial galactic nuclei) the star-SMBHB scattering process is efficient enough to avoid the "final parsec problem" and to let many SMBHB coalesce in a relatively short time\citep{yu02,mer04,ber06,khan11,pret11}.

Observational evidence for SMBHBs, in particular statistical data on their occurrence, could tell how efficiently they would shrink and coalesce. Currently only few candidates for SMBHB are known through indirect electromagnetic detections such as periodic variabilities, double-peaked emission line profiles and so on \citep{leht96, shen13, liu14, yan15}. Getting to sub-parsec separations in distant galaxies, especially in gas poor environments, is a considerable challenge for spatial resolution. But compact SMBHBs could be GW sources, which may be detected by the planned space based Laser Interferometer Space Antenna (LISA)\footnote{https://www.elisascience.org/} and ongoing Pulsar Timing Array (PTA) \citep{fost90, verb16}. However, there is still no detection from PTA even after the characteristic amplitude of the gravitational wave background has been constrained to $\sim 10^{-15}$ \citep{shan15,baba16,arzo16}. That could be caused by overestimation of either the GW characteristic amplitude of SMBHBs or the event rates \citep{shan15,sesa16}. Actually, how many SMBHBs could be detected by GW instruments is still uncertain. The detection rate depends on many unclear factors, such as the merger rate, the mass ratio distribution, the lifetime of the binary system, etc. In another word, it is susceptible to the evolution of SMBHBs. Therefore observations with statistical data on SMBHB before coalescence are needed to reveal the internal properties of the evolution of SMBHB.

In particular in gas poor environments some effective probes which can help us to investigate SMBHBs are needed. Tidal disruption events (TDEs) are potentially such a probe. A star can be torn up by the tidal force when it is closely encountering an SMBH. The critical distance for that to happen is defined as the tidal radius \citep{hil75,ree88,evan89,guil13}
\beq
\rt\backsimeq \mu\rstar(\Mbh/\mstar)^{1/3},
\label{eq:rt}
\eeq
where $\mu$ is a dimensionless parameter of order unity (depending on the stellar structure), $\rstar$, $\mstar$ and $\Mbh$ are the stellar radius, the stellar mass and the mass of BH, respectively. The bound fraction of the debris of the disrupted star forms an accretion disk around the dormant SMBH and temporarily illuminates it, accompanied by a flare from $\gamma$-ray to radio bands, which has been confirmed by many observations \citep[][and references therein]{geza13,komo15}. Similarly, there will be TDEs from SMBHBs too. Due to the gravitational perturbation from the companion BH, the TDE light curve of SMBHB may have characteristic drops, which has been confirmed by both theoretical and numerical studies \citep{liu09,coug17,vign18}, and a candidate also has been found in observation \citep{liu14}.

Though there are only tens of candidates and one candidate for TDEs in single SMBHs and in SMBHBs, respectively,
theoretical estimates of tidal disruption rates (TDR) exist for long time and by many authors, for both the single SMBH and the SMBHB. The theoretical TDR of a single SMBH in isolated spherical galaxies is $\sim 10^{-6}-10^{-4}$~$\yr^{-1}$ per galaxy, and can be enhanced a few times in flattened systems \citep{mago99,syer99,wan04,vasi13,ston16}. TDRs of SMBHB models, on the other hand, can be orders of magnitude lower if the binary is compact and has been embedded in a spherical isotropic galactic core dominated by two-body relaxation \citep{chen08}. However, the perturbation from the companion SMBH and the triaxial stellar distribution induced by the galaxy merger can actually enhance the TDR significantly \citep{ivan05,chen09,wegg11}. According to the theoretical analysis made by \citet{liu13}, in merging galaxies before the formation of a bound SMBHB, the TDR can be enhanced up to $\sim 10^{-2} \yr^{-1}$. However, during and after the binary formed, the interaction between stellar environment and the SMBHB is so rapid and presumably chaotic that both analytical estimations and scattering experiments are of limited value to reveal the underlying physical processes.

Rapid development of computing technology in hardware and software, notably the use of General Purpose Graphics Processing Units (GPGPU) for numerical computations in science, has promoted direct \nbody simulations of star clusters and galactic nuclei. This has led to many models based on direct \nbody simulations about TDR in spherical or axisymmetric nuclear stellar clusters around single SMBH\citep{bau04,zho14,zho15,pana18}. While numerical investigations also exist about the dynamical evolution of two SMBHs in merging galaxies \citep{ber06,khan11,pret11,khan18}, models studying the TDE evolution in merging galaxies and for SMBHB are very rare. \citet[][hereafter Paper I]{li17} has investigated the TDR evolution of two SMBHs and SMBHBs in equal mass mergers, and find that the TDR can be boosted significantly by up to two orders of magnitude during the phase when two SMBHs form a binary. But their results are only valid for equal mass mergers, which should be rare compared to the more common minor mergers \citep{fakh08,hopk10}. Just recently, \citet{pfis19} studied the TDR evolution in gaseous minor merger systems through hydrodynamical simulations, and found that the TDR of the secondary BH will be temporarily enhanced. However, their estimation of TDR is based on analytical models with spherical density distribution assumption, which is not the case when SMBHB is forming, and it can not resolve scattering between stars and SMBHB.

\citet{khan12} have shown that the so-called "final parsec problem" does not exist even for minor mergers, due to the merger induced triaxiality. But they find that compared with major mergers, it takes longer time to form a compact bound SMBHB in a minor merger system. We showed in Paper I for major mergers, that the TDR can be boosted significantly during the stage of the formation of a bound SMBHB (which we called "phase II" in Paper I), but only  for a very short period of time. Hence, in total the phase II in our previous work only contributed a quarter of all TDEs in the entire evolution. In this paper we want to study the TDR in minor mergers, which have not been considered in the previously cited papers. Minor mergers are more frequent, and there is evidence from previous work that the "phase II" with enhanced TDR may last longer. Therefore we speculate at this stage that minor mergers could contribute a significant fraction of TDEs in total, comparable or even more than for major mergers. So, in this paper, we are focusing on unequal mass mergers.


This paper is organized as follows. In Section~\ref{mtd} we describe our simulation method and models. The result for the evolution of disruption rate are given in Section~\ref{res}. In Section~\ref{theory}, we briefly introduce how to estimation disruption rate analytically in unequal mass mergers. Then we have a discussion about how to extrapolate our simulation results to a few typical realistic minor mergers, and provide a short discussion about the implications to observations in Section~\ref{dis}. A short summary is allocated to Section~\ref{sum}.

\section{Numerical Simulations}
\label{mtd}

All of stars which can be disrupted by the SMBH will have orbits inside a cone region in the phase space, which is the so-called "loss cone". In principal, how many stars are inside loss cone and how fast stars outside loss cone can be scattered inwards will directly decide the TDR. Therefor, as demonstrated in Paper I, since different phases have different dominated loss cone refilling mechanisms, the TDR evolution in equal mass mergers can be estimated by combining analytical models with numerical \nbody simulations. Similarly, this strategy also works for minor mergers. Here we will briefly introduce it. More details can be found in Paper I.

Our model assumes that there are two unequal mass galaxies or nuclei, and each of them has an SMBH in the center. Here $\Mbhh$ and $\Mbhl$ represent the mass of the primary and the secondary SMBH respectively, $N_1$ and $N_2$ represent the number of stars in the larger and the smaller galaxy/nucleus, respectively, and $q = \Mbhl / \Mbhh$ is the mass ratio between two SMBHs. For simplicity, we assume that all the stars in merging galaxies have the same mass, and the mass ratio between a SMBH and its host galaxy/nucleus $f = \Mbh / M_{\rm G} = 0.01$ is the same for both galaxies/nuclei, which namely the mass ratio of two galaxies/nuclei is $M_{\rm G2}/M_{\rm G1} = N_2/N_1 = \Mbhl/\Mbhh = q$. In our models, we fix the mass of the primary galaxies/SMBHs, and only vary the mass of the secondary galaxies/SMBHs.

For simplicity, a spherical Dehnen model has been adopted to represent the stellar distribution of each galaxies/nuclei \citep{deh93}. And the density profile of this Dehnen model is
\beq \rho(r)=\frac{3-\gamma}{4\pi}\frac{Ma}{r^\gamma(r+a)^{4-\gamma}} ,
\label{Dehnen rho}
\eeq
where $a$, $M$ and $\gamma$ denote the scaling radius, the total mass of galaxy/nucleus and the density profile index, respectively. In accordance with Paper I, we adopt the units $G=M=M_{\rm G1}=a=1$ hereafter. Similar to Paper I, the relation between numerical and physical quantities can be derived as
\begin{eqnarray}
[T] &=& \left(\frac{GM}{a^3}\right)^{-1/2} \nonumber \\
    &=& 1.491\times 10^7(2^{\frac{1}{3-\gamma}}-1)^{3/2}\left( \frac{M}{10^{9}\msun}\right)^{-1/2}\left(\frac{r_{1/2}}{1\kpc}\right)^{3/2} \yr, \\
    \label{eq:scalingT}
[V] &=& \left(\frac{GM}{a}\right)^{1/2} \nonumber \\
    &=& 65.58\times (2^{\frac{1}{3-\gamma}}-1)^{-1/2}\left( \frac{M}{10^9\msun}\right)^{1/2}\left(\frac{r_{1/2}}{1\kpc}\right)^{-1/2} \kms, \\
    \label{eq:scalingV}
[R] &=& a=(2^{\frac{1}{3-\gamma}}-1)\left(\frac{r_{1/2}}{1\kpc}\right) \kpc, \\
    \label{eq:scalingr}
[\dot{M}] &=& M / [T] \nonumber \\
          &=& 67.07\times(2^{\frac{1}{3-\gamma}}-1)^{-3/2}\left( \frac{M}{10^9\msun}\right)^{3/2}\left(\frac{r_{1/2}}{1\kpc}\right)^{-3/2} \msun/\yr.
          \label{eq:scalingMdot}
\label{eq:scalingr1}
\end{eqnarray}

Since we want to focus on how TDR evolves in unequal mass mergers with different mass ratio, here we fix the density profile index to $\gamma = 1.0$. We select a fiducial model with $M_{\rm G1} = 4\times 10^{9} \msun$, $\Mbhh = 4\times 10^{7} \msun$, half mass radius of primary galaxy/nucleus $r_{1/2} = 1 \kpc$ and $q = 0.25$.

Due to the huge number of stars in a typical galaxy, even with the most powerful computer in the world, it is still impossible to make an one-to-one direct \nbody simulation. The maximum particle number can be managed with reasonable cost nowadays is only a few million. That means we have to use millions of particles to simulate a galaxy, and then make extrapolations. With this limited particle resolution, there are some artificial effects should be taking into account. For instance, the two-body relaxation time scale $T_{\rm r}$ is roughly proportional to the particle number. Smaller particle numbers will result in artificially shorter two-body relaxation time scale, which can not be avoid in direct \nbody simulations. Besides, the mass ratio between the SMBH and solar type star should be approximate $10^6 - 10^9$ in real galaxies. While in direct \nbody simulations, if $f = 0.001$, and $q = 0.1$, with one million particle number for the primary galaxy/nucleus, the mass ratio of secondary SMBH to stellar particles should be only $100$. With this small mass ratio, stellar particles can make significantly enlarged gravitational influences to the BH than the situation in typical galaxies. In order to reduce those influences we have to set the particle number and BH to star mass ratio as large as we can. As a result, the particle number of the primary galaxy/nucleus has been set to one million in most of our models. And the mass ratio between SMBH and host galaxy has been fixed to $f = 0.01$, which is relatively larger than the popular ratio $\sim 10^{-3}$ based on observation. However, it is still reasonable in many ellipticals \citep{korm13}. With this configuration, and taking into account that the total stellar mass of the primary galaxy is fixed to one in our model, the BH to star mass ratio in simulation can achieve to $10^4$ for the primary SMBH, and $10^3$ for the secondary SMBH in minor mergers with the largest disparities. Despite of two or three orders of magnitude lower than the ratio between a $10^6\msun$ SMBH and a solar type star, it is already close to the reality. And this is the highest particle resolution we can achieve with reasonable computation resources nowadays.

In addition to that, smaller particle numbers will also lead to less TDEs in the simulation, which makes the statistical investigation has larger uncertainties. In our fiducial model, we have $[T] \sim 3.97\Myr$, $[V] \sim 102\kms$, $[R] \sim 0.4\kpc$ and $[\dot{M}] \sim 252 \msun/\yr$. Thus the tidal radius of a solar type star should be $\sim 10^{-8}$. That means, even with millions of particles adopted, we still do not have enough particle resolution to collect enough TDEs for statistical analyze with this tiny tidal radius.

In conclusion, a compromised solution is to adopt a set of simulations with different larger tidal radii and limited particle numbers which can be managed by the direct \nbody integration, and then extrapolate to the condition close to real galaxies. Since this scheme has been adopted in many literatures \citep[][and references therein]{bau04,broc11,li12,zho14}, we will not get into details here. Interested readers are referred to Paper I with more discussion.

Similar to Paper I, as the first step, we put an SMBH into the center of each galaxy/nucleus and make them dynamically relaxed. Then two nuclei are set with separation $d \sim 20$, and make sure that they have parabolic orbits with the first pericenter $d \sim 1$. We adopt a direct \nbody code $\varphi$\,-{\sc Grape}/$\varphi$\,-GPU for integration. It is an accurate and efficient parallel code with fourth-order Hermite integrator and accelerated by GPU, which is the same code adopted in Paper I \citep{maki92,ber05,harf07}. All integrations are calculated on the $laohu$ GPU cluster in National Astronomical Observatories of China (NAOC).

Table~\ref{tab:para} lists all the models we have integrated. The column (1) to (4) are, respectively, the model index, the mass ratio, the particle number of primary nucleus, and the tidal radius. The main simulations are concerning on different mass ratios from model Q10 to Q1. And there are two sets of models focusing on the influence of different particle numbers and tidal radii with model N250, N500 and RT5-5 to RT1-3, respectively. Model Q4 is our fiducial model.

\begin{deluxetable}{cccc}
    \tablewidth{0pt}
    \tabletypesize{\scriptsize}
    \tablecaption{Parameters of simulation models \label{tab:para}}
    \tablehead{
    \colhead{$Model \, No.$} &
    \colhead{$q$} &
    \colhead{$N_{\rm 1}$} &
    \colhead{$r_{\rm t}$} \\
    \colhead{(1)} &
    \colhead{(2)} &
    \colhead{(3)} &
    \colhead{(4)}
    }
    \startdata
Q10 & 1:10  & $1\rm M$ & $5\times 10^{-4}$  \\
Q8 & 1:8  & $1\rm M$ & $5\times 10^{-4}$  \\
Q5 & 1:5  & $1\rm M$ & $5\times 10^{-4}$  \\
Q4 & 1:4  & $1\rm M$ & $5\times 10^{-4}$  \\
Q3 & 1:3  & $1\rm M$ & $5\times 10^{-4}$  \\
Q2 & 1:2  & $1\rm M$ & $5\times 10^{-4}$  \\
Q1 & 1:1  & $1\rm M$ & $5\times 10^{-4}$  \\
\\
\hline \\
N250 & 1:4 & $250\rm K$ & $5\times 10^{-4}$  \\
N500 & 1:4 & $500\rm K$ & $5\times 10^{-4}$  \\
\\
\hline \\
RT5-5 & 1:4 & $1\rm M$ & $5\times 10^{-5}$  \\
RT1-4 & 1:4 & $1\rm M$ & $1\times 10^{-4}$  \\
RT1-3 & 1:4 & $1\rm M$ & $1\times 10^{-3}$  \\
    \enddata

\tablecomments{ Col.(1): Model sequence number. Col.(2): Mass ratio of two galaxies/SMBHs. Col.(3): Particle numbers of primary galaxy/nucleus adopted in calculations. Col.(4): tidal radius $\rt$. For reference, we have two additional models N500 and N250, with particle number of primary galaxy/nucleus $N=500{\rm K}$ and $N=250{\rm K}$ respectively. And there are three models RT5-5 to RT1-3 with variable tidal radius.}
\end{deluxetable}

\section{Results}
\label{res}

In this section, we will present our numerical simulation results. As mentioned in Paper I and Appendix~\ref{sec:app}, the entire evolution of two SMBHs in merging galaxies can be roughly divided into three phases, which corresponds to two SMBHs are far away from each other, forming a bound binary, and evolving to a hard binary, respectively. And the evolution of TDR in each phase is dominated by different mechanism. However, how to divide different phases is an ambiguous concept without accurate criteria. In order to make detailed comparison among different models, a precise criteria in our numerical models is needed. Here we follow the same strategy adopted in Paper I to empirically divide the evolution into three phases. The criteria are based on the fluctuation of the TDE number rate relative to the averaged rate during the entire evolution. Here we define $\Delta N_{\rm TDE}$ as the variation between the peak rate and the averaged rate. The start/end point of phase II corresponds to the time when the number of TDEs rapidly increase/decrease a large fraction of $\Delta N_{\rm TDE}$. In this work we empirically fix the fraction to $0.4$, to prevent misjudgement in some extreme conditions. However, it is not a fine tuned fraction. Slightly changes on this value will not make significant influences to results. In short, the criteria is based on the TDR evolution. We have also tried to divide the three phases according to the dynamical evolution of two SMBHs, which means the separation evolution of two SMBHs, and found they are roughly consistent. In principle, the latter is the most physical criteria. However, in practical the separation of two BHs are varying rapidly, which often gives wrong judgment. Nevertheless, in most of cases, the phase II we have divided in our models are roughly corresponding to $1 - 0.05$ influence radius of the SMBHB.

\begin{figure}
\begin{center}
\includegraphics[width=0.8\textwidth,angle=0.]{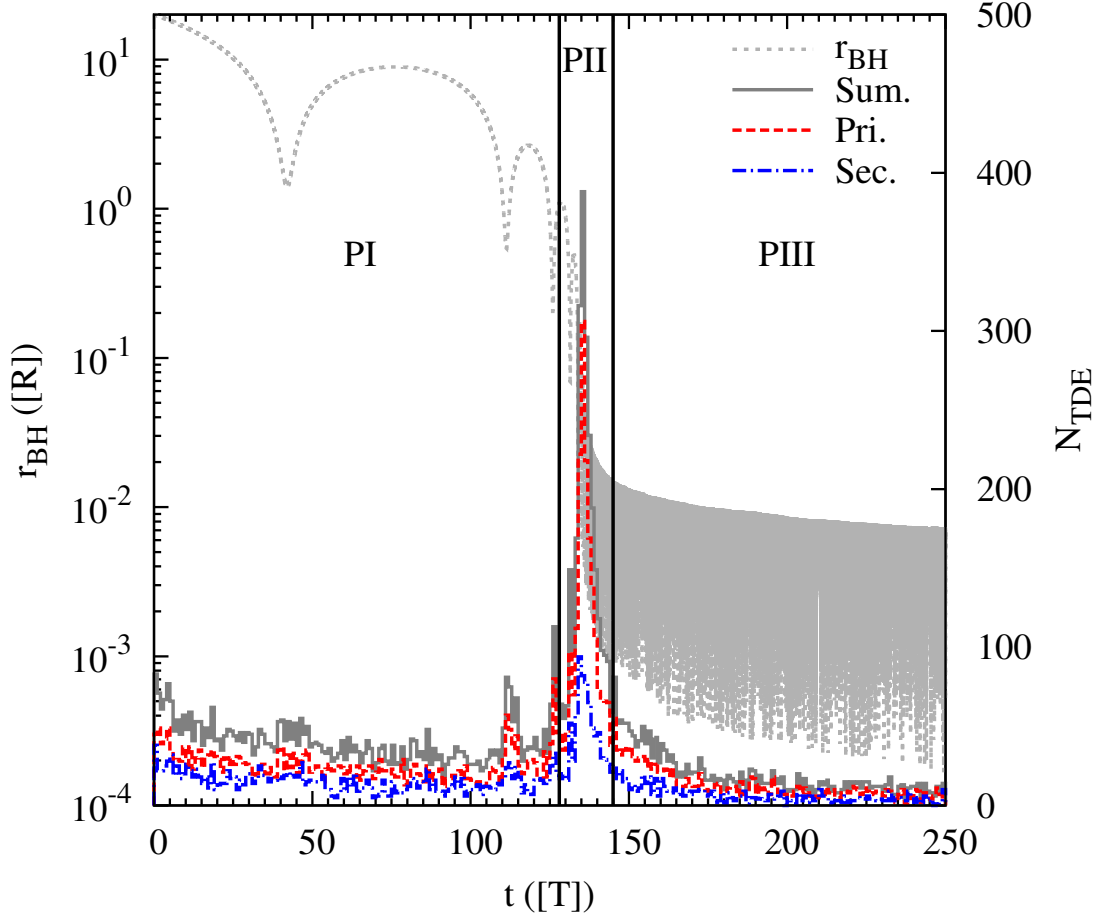}
\caption {Tidal disruption evolution of the fiducial Model Q4. Here the \textit{x}-axis and the left \textit{y}-axis represent evolved time and $\rbh$, the separation of two SMBHs, respectively. The right \textit{y}-axis gives corresponding tidal disruption number count $N_{\rm TDE}$ within one \nbody time unit bin. Grey dotted and solid lines represent $\rbh$ and total $N_{\rm TDE}$ of two SMBHs respectively. Red dashed and blue dash dotted lines denote $N_{\rm TDE}$ of the primary and the secondary SMBH, respectively. Two vertical solid lines divide the evolution into three phases. (A color version is available in the online journal.)
\label{fig:TD-std}}
\end{center}
\end{figure}

Fig.~\ref{fig:TD-std} demonstrates the tidal disruption evolution of our fiducial mode Q4. The grey dotted line represents the separation of two SMBHs, and the grey solid line represents the tidal disruption event number count $N_{\rm TDE}$ within one \nbody time unit bin for the summation of two SMBHs. Red dashed and blue dash dotted lines are, respectively, the $N_{\rm TDE}$ of the primary SMBH and the secondary SMBH. Similar to equal mass mergers in Paper I, the evolution of unequal mass mergers can be divided into three phases, which has been separated by two vertical solid lines. And it is obvious that our criteria successfully picks up the phase II, when the bound SMBHB is forming and the TDR is achieving to the maximum value.

According to Fig.~\ref{fig:TD-std}, there are roughly constant disruption rates achieved in phase I and III, while a significant peak in phase II. As mentioned in Paper I, two SMBHs and the star clusters around them can strongly perturb surrounding stars in phase II, which leads to very efficient loss cone refilling. As a result, there is a significant tidal disruption burst in phase II. Due to the same reason, there are also several smaller bursts corresponding to pericenters in phase I. Besides, the TDR of the primary SMBH is obviously higher than the rate of the secondary SMBH. Specially in phase II, the primary SMBH dominates the contribution.

\begin{figure}
\begin{center}
\includegraphics[width=0.8\textwidth,angle=0.]{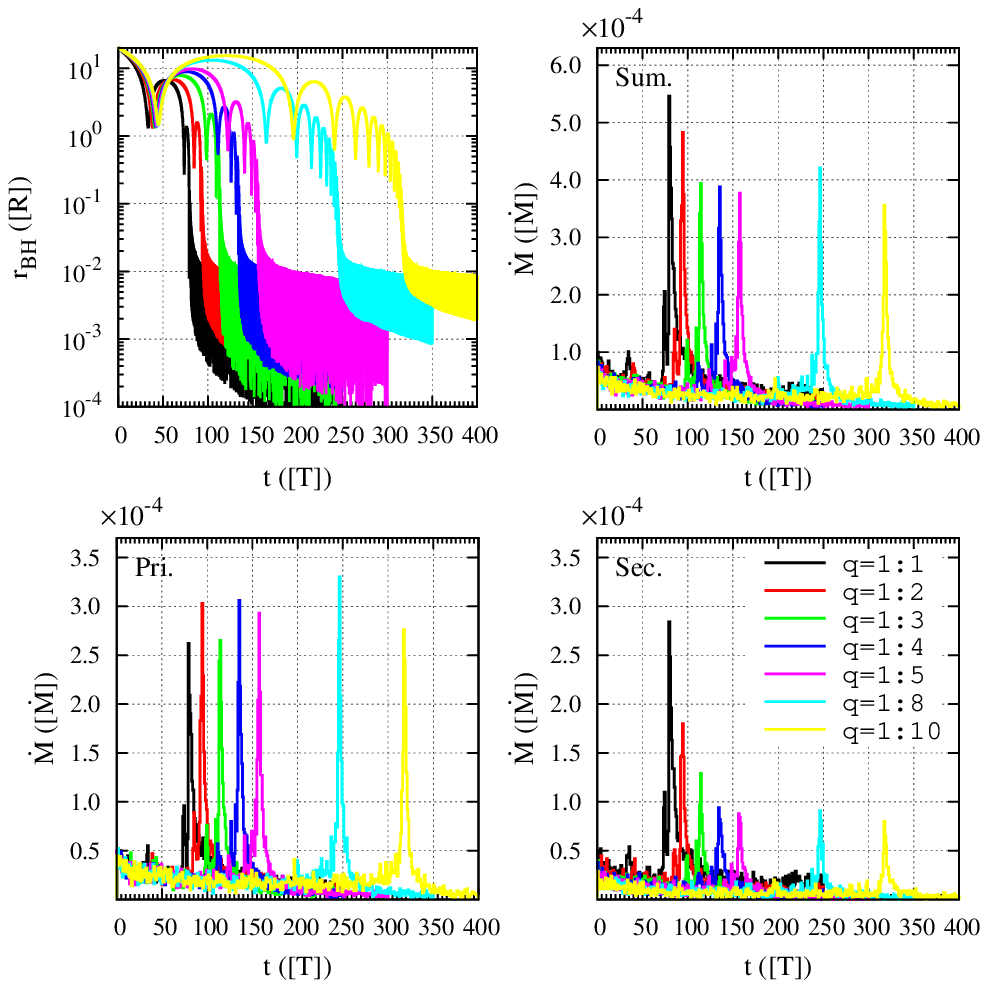}
\caption {Orbital evolution of SMBHBs and corresponding tidal disruption accretion rates. Here the \textit{x}-axis denotes the evolved time. The \textit{y}-axis represents SMBH separation $\rbh$ in the top left panel, and tidal disruption accretion rates in other three panels. The top left panel shows the evolution of $\rbh$ for Model Q10 to Q1. The top right, bottom left and bottom right panels are corresponding tidal disruption accretion rates for the summation of two SMBHs, the primary SMBH, and the secondary SMBH, respectively.
\label{fig:TD-all}}
\end{center}
\end{figure}

Similar results can be found in all of the rest models. Fig.~\ref{fig:TD-all} gives the evolution of the separation and disruption rate for all models from Q10 to Q1. As shown in the top left panel, all of the models have similar separation evolution for SMBHBs. But the lighter the secondary SMBH is, the longer dynamical evolution it has. Because a more massive SMBH usually experiences stronger influence from dynamical friction, which makes two SMBHs could more efficiently sink to the galactic center. This effect is more significant for models with very small mass ratios, for instance, model Q10 and Q8. In order to finish the integration of such kind of systems and achieve enough data for statistical analyze in phase III, we have to extend simulation time. For simplicity, we do not consider inspiral orbits here, which is too time consuming for small mass ratio models. We set the termination time for Model Q4, Q3, Q2 and Q1 to $t = 250$, and extend it to $t = 400$, $t = 350$ and $t = 300$ for model Q10, Q8 and Q5, respectively.

The rest three panels in Fig.~\ref{fig:TD-all} represent the evolution of tidal disruption accretion rate for the summation of two BHs, the primary SMBH and the secondary SMBH, respectively. According to the bottom two panels, the peak rate of the primary SMBH does not significantly depend on $q$, while the secondary SMBH is opposite. Actually, as shown by Equation~(\ref{eq:TDRsPIIM2}) in the following analysis, the disruption rate of the secondary SMBH depends on the mass of itself in phase II. While the rate of the primary SMBH does not sensitively depend on $q$, which is consistent with the results here.

\begin{figure}
\begin{center}
\includegraphics[width=0.8\textwidth,angle=0.]{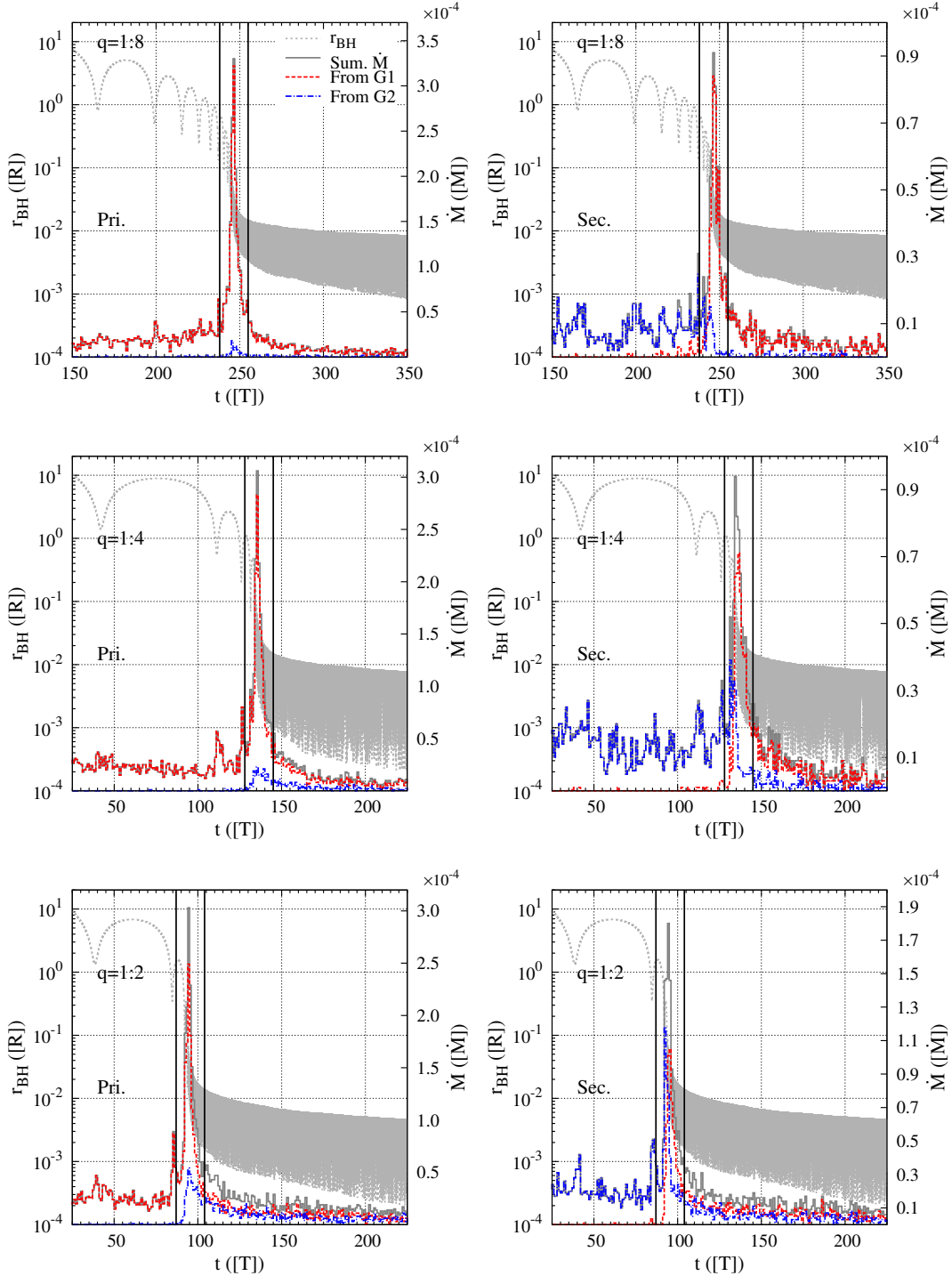}
\caption {How stars from different galaxies/nuclei contribute to tidal disruption of the primary and the secondary SMBH. Similar to Fig.~\ref{fig:TD-std}, the \textit{x}-axis and the left \textit{y}-axis represent evolved time and $\rbh$, respectively. The right \textit{y}-axis gives corresponding tidal disruption accretion rate. Grey dotted and solid lines represent $\rbh$ and the total accretion rate from both galaxies/nuclei. Red dashed and blue dash dotted lines denote the contribution from the primary and the secondary galaxy/nucleus, respectively. The top, middle and bottom panels are the results of mode Q8, Q4 and Q2, respectively. The left and right columns are results of the primary and the secondary SMBHs, respectively. Similar to Fig.~\ref{fig:TD-std}, two vertical solid lines in each penal divide the evolution into three phases. (A color version is available in the online journal.)
\label{fig:TD-G1G2}}
\end{center}
\end{figure}

Since the mass and particle number of two galaxies/nuclei are unequal, it is interesting to know how stars from different galaxies/nuclei contribute to the tidal disruption. Fig.~\ref{fig:TD-G1G2} presents the similar evolution results as Fig.~\ref{fig:TD-std} for model Q8, Q4 and Q2, which corresponds to the top, middle and bottom panels respectively. Here we separately record tidal disruption of two SMBHs, which gives results of the primary SMBH and the secondary SMBH in left and right columns individually. In addition to that, we also distinguish those disrupted stars according to their original host galaxies/nuclei. For reference, the separation of two SMBHs is marked as dotted grey line and the total rate taking into account stars from both galaxies/nuclei is marked as grey solid line. The red dashed and blue dash dotted lines are the contribution from the primary and the secondary galaxy/nucleus, respectively. As shown in the figure, the TDRs of two SMBHs are dominated by their own host galaxies in phase I for all of the mass ratios, which means their tidal disruption evolution are more or less the same as the situation in an isolated system. While in phase II when two SMBHs are close enough to form a binary, the contribution from the companion becomes more and more significant. For the primary SMBH, although most of disrupted stars come from the primary galaxy/nucleus for all models, there are still some contributions from the second galaxy/nucleus. The more massive the secondary SMBH is, the more stars from the secondary galaxy/nucleus disrupted by the primary SMBH. The results for the secondary SMBH are more interesting. We find that the rates of the secondary SMBH in phase II and III are actually dominated by stars originated from the other galaxy/nucleus in minor mergers.

In model Q8 and Q4, the contribution of stars from the secondary galaxy/nucleus will sharply drop in phase II. While it does not appear in model Q2, which corresponds to a major merger. Though there are only three models have been shown in Fig.~\ref{fig:TD-G1G2}, we actually have checked all of other models. This effect is significant for all models with mass ratio $q \lesssim 0.3$. After the formation of the SMBHB in minor mergers, when the binary evolving to phase III, the disruption rate will be dominated by stars from the primary galaxy/nucleus. While there are still significant contribution from the secondary galaxy/nucleus in major mergers.

These results indicate that different mechanisms dominate the loss cone refilling in different phases. In phase I, since two SMBHs are far away from each other, the tidal disruption will be dominated by stars in their own host galaxies. While in phase II, when the secondary SMBH getting closer to the primary SMBH, more and more stars around the former will be tidally striped. If the secondary SMBH is significantly light than the primary SMBH, most of its stars will be finally striped. As a result, the loss cone refilling of the secondary SMBH in the late stage of phase II will be like a scatter process, which means the tidal disruption accretion rate should be proportional to the cross-section. Thus we find in minor mergers, stars disrupted by the secondary SMBH are mainly from the secondary galaxy/nucleus in phase I, and then suddenly switch to the other galaxy in phase II and III. Understanding this transition is important for our later derivation of the extrapolation relations.

With divided three phases we can calculate corresponding averaged tidal disruption accretion rate of every model. Table~\ref{tab:res-time} lists the demarcation points of different phases, and also the time $t_{\rm p}$ corresponding to the peak rate. In order to simplify the comparison, we assume that the duration of phase III is $70$ time units, and the phase I starts from $t = 20$ to avoid some numerical artificial effects at the beginning of the integration. Corresponding results can be found in Table~\ref{tab:res-all}. The first column in the table is model index, and the next three columns, Col. (2) to (4), are the averaged rates of phase I for the summation of two SMBHs, the primary SMBHs and the secondary SMBHs respectively. The next three columns, Col. (5) to (7) are the rates of phase II and the last three columns are the rates of phase III.

\begin{deluxetable}{cccccccccccc}
    \tablewidth{0pt}
    \tabletypesize{\scriptsize}
    \tablecaption{Demarcation points for different phases. \label{tab:res-time}}
    \tablehead{
    \colhead{Model} &
    \colhead{$t_{\rm PI-PII}$} &
    \colhead{$t_{\rm PII-PIII}$} &
    \colhead{$t_{\rm end}$} &
    \colhead{$t_{\rm p}$}\\
    \colhead{(1)} &
    \colhead{(2)} &
    \colhead{(3)} &
    \colhead{(4)} &
    \colhead{(5)}
    }

    \startdata
Q10 & 310 & 327 & 397 & 317.5  \\
Q8 & 238 & 255 & 325 & 246.5  \\
Q5 & 149 & 166 & 236 & 157.5  \\
Q4 & 128 & 145 & 215 & 135.5  \\
Q3 & 106 & 123 & 193 & 114.5  \\
Q2 & 87  & 104 & 174 &  94.5  \\
Q1 & 73  & 92  & 162 &  79.5  \\
N250 & 123 & 139 & 209 & 130.5  \\
N500 & 126 & 142 & 212 & 133.5  \\
RT5-5 & 129 & 144 & 214 & 136.5  \\
RT1-4 & 125 & 142 & 212 & 134.5  \\
RT1-3 & 127 & 142 & 212 & 133.5  \\
    \enddata

\tablecomments{ Col.(1): Model index. Col.(2): Demarcation point of phase I and II. Col.(3): Demarcation point of phase II and III. Col.(4): The end of phase III for TDR calculation. Col.(5): The time corresponding to maximum total TDR. }
\end{deluxetable}

\begin{deluxetable}{cccccccccccc}
    \tablewidth{0pt}
    \tabletypesize{\scriptsize}
    \tablecaption{Tidal disruption accretion rates in different phases. \label{tab:res-all}}
    \tablehead{
    \colhead{Model} &
    \colhead{$\dot{M}_{\rm PI}$} &
    \colhead{$\dot{M}_{\rm 1PI}$} &
    \colhead{$\dot{M}_{\rm 2PI}$} &
    \colhead{$\dot{M}_{\rm PII}$} &
    \colhead{$\dot{M}_{\rm 1PII}$} &
    \colhead{$\dot{M}_{\rm 2PII}$} &
    \colhead{$\dot{M}_{\rm PIII}$} &
    \colhead{$\dot{M}_{\rm 1PIII}$} &
    \colhead{$\dot{M}_{\rm 2PIII}$} \\
    \colhead{} &
    \colhead{$\times 10^{-5}$} &
    \colhead{$\times 10^{-5}$} &
    \colhead{$\times 10^{-5}$} &
    \colhead{$\times 10^{-4}$} &
    \colhead{$\times 10^{-4}$} &
    \colhead{$\times 10^{-4}$} &
    \colhead{$\times 10^{-5}$} &
    \colhead{$\times 10^{-5}$} &
    \colhead{$\times 10^{-5}$} \\
    \colhead{(1)} &
    \colhead{(2)} &
    \colhead{(3)} &
    \colhead{(4)} &
    \colhead{(5)} &
    \colhead{(6)} &
    \colhead{(7)} &
    \colhead{(8)} &
    \colhead{(9)} &
    \colhead{(10)} \\
    }

    \startdata
Q10 & $2.66$ & $1.93$ & $0.73$ & $1.30$ & $1.00$ & $0.30$ & $1.55$ & $1.03$ & $0.51$  \\
Q8 & $2.97$ & $2.06$ & $0.91$ & $1.44$ & $1.09$ & $0.35$ & $1.74$ & $1.18$ & $0.56$  \\
Q5 & $3.51$ & $2.35$ & $1.16$ & $1.46$ & $1.07$ & $0.39$ & $2.07$ & $1.37$ & $0.70$  \\
Q4 & $3.86$ & $2.43$ & $1.43$ & $1.55$ & $1.10$ & $0.44$ & $2.36$ & $1.57$ & $0.79$  \\
Q3 & $4.21$ & $2.52$ & $1.69$ & $1.52$ & $1.02$ & $0.49$ & $2.60$ & $1.70$ & $0.90$  \\
Q2 & $4.79$ & $2.71$ & $2.09$ & $1.84$ & $1.12$ & $0.73$ & $3.99$ & $2.43$ & $1.56$  \\
Q1 & $5.22$ & $2.66$ & $2.56$ & $2.16$ & $1.04$ & $1.12$ & $5.77$ & $2.91$ & $2.86$  \\
N250 & $7.26$ & $4.84$ & $2.42$ & $1.66$ & $1.24$ & $0.42$ & $2.76$ & $1.75$ & $1.01$  \\
N500 & $5.64$ & $3.70$ & $1.94$ & $1.70$ & $1.19$ & $0.51$ & $2.54$ & $1.75$ & $0.79$  \\
RT5-5 & $1.40$ & $0.89$ & $0.52$ & $0.29$ & $0.20$ & $0.09$ & $0.45$ & $0.30$ & $0.11$  \\
RT1-4 & $1.96$ & $1.22$ & $0.74$ & $0.47$ & $0.34$ & $0.13$ & $0.75$ & $0.52$ & $0.23$  \\
RT1-3 & $5.34$ & $3.36$ & $1.98$ & $2.68$ & $1.84$ & $0.83$ & $3.59$ & $2.27$ & $1.32$  \\
    \enddata

\tablecomments{ Col.(1): Model index. Col.(2): Averaged total accretion rate during phase I. Col.(3): Averaged rate of the primary SMBH in phase I. Col.(4): Averaged rate of the secondary SMBH in phase I. Col.(5): Averaged total accretion rate in phase II. Col.(6): Averaged rate of the primary SMBH in phase II. Col.(7): Averaged rate of the secondary SMBH in phase II. Col.(8): Averaged total accretion rate in phase III. Col.(9): Averaged rate of the primary SMBH in phase III. Col.(10): Averaged rate of the secondary SMBH in phase III. }
\end{deluxetable}

\begin{figure}
\begin{center}
\includegraphics[width=0.8\textwidth,angle=0.]{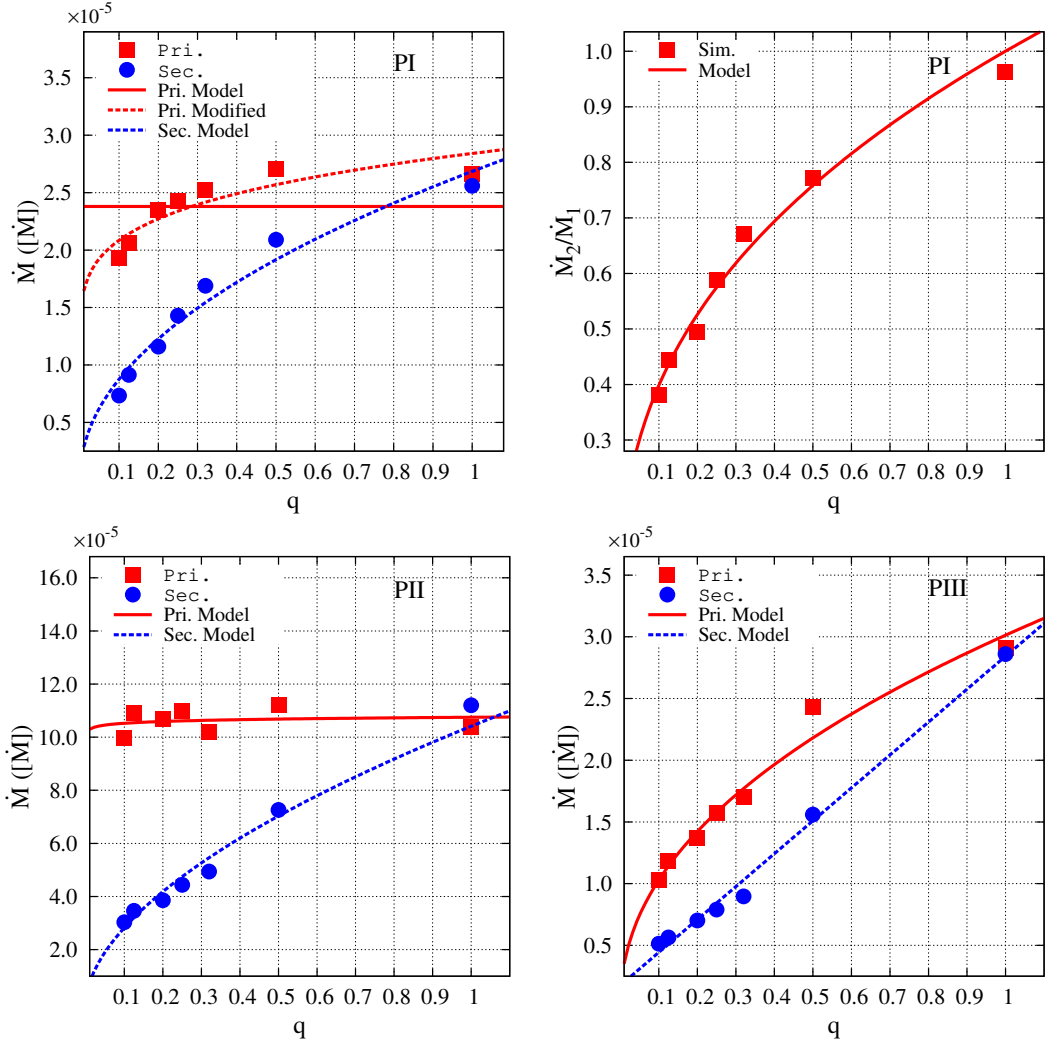}
\caption {Phase averaged tidal disruption accretion rates for different mass ratios. Top left, bottom left and bottom right panels are, respectively, the averaged rates of two SMBHs in phase I, II and III. The \textit{y}-axis in these three panels denote accretion rate. Top right panel is the rate ratio between the secondary and the primary SMBH in phase I, which is represented by the \textit{y}-axis. Red filled squares and blue filled circles are from \nbody simulation results for the primary and the secondary SMBHs, respectively. Red solid and blue dashed lines are fitting results based on our models for the primary and the secondary SMBHs, respectively. Red dashed line in top left panel is the fitting result based on the model with pericenter perturbation effect considered. (A color version is available in the online journal.)
\label{fig:qdep}}
\end{center}
\end{figure}

Fig.~\ref{fig:qdep} shows how the tidal disruption accretion rate depends on the mass ratio. The top left, bottom left and bottom right panels are results for phase I, II and III, respectively. The top right panel is the dependence of the accretion rate ratio on the mass ratio in phase I. Here the red solid squares and blue solid circles are averaged results based on simulations of model Q10 to Q1. The red solid lines and blue dashed lines are fitting results based on the analytical predictions in Section~\ref{theory}.

In phase I, since two SMBHs are relatively far away from each other, their TDR should not be significantly perturbed by their companions. Thus the primary SMBH should keep a nearly constant TDR for different mass ratio. While the simulation results indicate a slightly increasing trend. We will discuss about it in Section~\ref{dis-PI}. On the other hand, since larger mass ratio corresponds to heavier secondary SMBH and higher TDR, the secondary SMBH should have a monotonically increasing rate, which is consistent with the simulation results in Fig.~\ref{fig:qdep}. With the assumption that the loss cone is almost full in phase II, the TDR of two SMBHs will be dominated by the perturbations from each other. As a result, the TDRs of both SMBHs should be increasing for larger mass ratio models. As demonstrated by the bottom left panel of Fig.~\ref{fig:qdep}, the results of the secondary SMBH are well consistent with that prediction, and the primary SMBH also shows a slightly increasing relation. In the last phase, as presented by the bottom right panel of Fig.~\ref{fig:qdep}, TDRs of both SMBHs are monotonically increasing related to the mass ratio. In Section~\ref{Aly-PIII} and \ref{dis-extr} we have carefully analysis and give more detailed explanations and fitting formulae to the results here.

\section{Analytical model}
\label{theory}

In Paper I, by assuming that the loss cone refilling are dominated by two-body relaxation and perturbation in phase I and II, respectively, the TDR in equal mass mergers can be estimated analytically. Most of their assumptions and results are still valid for minor mergers. In this section we will derive the TDR of minor mergers based on some basic assumptions and results in Paper I, where more details can be found in Appendix~\ref{sec:app}.

\subsection{Phase I}
\label{Aly-PI}

With the assumption that the loss cone refilling is dominated by two-body relaxation in phase I, we can easily derive the accretion rate for both SMBHs. In Paper I, the mass accretion rate of the SMBH due to tidal disruption can be estimated by
\beq
\ln\dot{M_1}\approx \ln A - \frac{2\gamma-1}{8-2\gamma}\ln\left(\frac{N_1}{\ln\Lambda}\right) + \frac{9-4\gamma}{8-2\gamma}\ln\rt.
\label{eq:lnTDRsI}
\eeq
The definition of $A$ and the Coulomb logarithm $\ln\Lambda$ can be found in Appendix~\ref{sec:appPI}. Similarly, with $\Mbhl = \Mbhh q$ and $N_2 = N_1 q$, the mass accretion rate of the secondary SMBH can be derived as
\beq
\ln\dot{M_2}\approx \ln A - \frac{2\gamma-1}{8-2\gamma}\ln N_1 + \frac{7-7\gamma}{8-2\gamma}\ln q + \frac{2\gamma-1}{8-2\gamma}\ln(\ln \Lambda + \ln q) + \frac{9-4\gamma}{8-2\gamma}\ln\rt.
\label{eq:lnTDRs_2nd}
\eeq
As a result,
\beq
\frac{\dot{M_2}}{\dot{M_1}} = \left(1+\frac{\ln q}{\ln \Lambda}\right)^{\frac{2\gamma-1}{8-2\gamma}}q^{\frac{7-7\gamma}{8-2\gamma}}.
\label{eq:MdotR}
\eeq

As indicated by Equation~(\ref{eq:lnTDRsI}), the rate of the primary SMBH does not depend on $q$. The averaged rate hereafter should be constant.

\subsection{Phase II}
\label{Aly-PII}

In phase II, as demonstrated in Paper I and Appendix~\ref{sec:appPII}, we can consider the secondary SMBH as the perturber, which induced the replenishing process of the loss cone. Thus according to the assumption that the mass of the perturber $M_{\rm p} \propto \Mbhh q$, and Equations~(\ref{eq:TDRsPII}), we have
\beq
\dot{M}_1 \propto q^{\frac{4-2\gamma}{11}}\rt^{\frac{12-2\gamma}{11}}.
\label{eq:TDRsPIIM1}
\eeq

Analogously, the rate of the secondary SMBH in phase II is dominated by the perturbation from the primary BH and stars around it. Since $\Mbhl=\Mbhh q$, it can be derived as
\beq
\dot{M_2} \propto q^{\frac{7-3\gamma}{11}}\rt^{\frac{12-2\gamma}{11}}.
\label{eq:TDRsPIIM2a}
\eeq
It should be aware that the $\rt$ in Equations~(\ref{eq:TDRsPIIM1}) and (\ref{eq:TDRsPIIM2a}) is the corresponding tidal radius of each BH.

In order to get the results above, it assumes that most of stars disrupted by the secondary BH are from inner region with $r \lesssim \rinf$, and they are relatively far away from the perturber, which may not be the case for entire phase II. Our simulation results in Fig.~\ref{fig:TD-G1G2} show that, for minor mergers, the TDEs of the secondary SMBH in phase II are dominated by stars from its host galaxy at the beginning and then switch to stars from the other galaxy. It can be explained by the follows. At the beginning of phase II, the separation of two SMBHs is significantly larger than the influence radius of the secondary SMBH. The disruption should be mainly contributed by perturbed stars originally bound to the secondary SMBH. While at the end of phase II, the separation of two SMBHs has been shrunk rapidly, which is much more smaller than the influence radii of both BHs. Most of bound stars around the secondary BH have been stripped away during this process. Thus stars outside the influence radius could significantly contribute to the disruption rate.

Actually, for a minor merger, the secondary SMBH in the late stage of phase II can be considered as embedded in a stellar background which is bound to the primary BH. The tidal disruption accretion rate for such kind of systems will be proportional to the cross-section, which gives $\dot{M_2} \propto \Mbhl \rt \propto q \rt$ \citep{li12}. On average, we can assume that the average rate of the secondary SMBH in phase II follows
\beq
\dot{M_2} \propto \Mbhl^\alpha \rt^\beta \propto q^\alpha\rt^\beta,
\label{eq:TDRsPIIM2}
\eeq
where $\alpha$ and $\beta$ are undetermined indices. This is consistent with the results of \citet{mer04} with similar triaxial but none evolving stellar distribution, which found that $\dot{M}\propto \rt$ for $\gamma = 1$ and $\dot{M}\propto \rt^{0.5}$ for $\gamma = 2$.

According to the discussion in Paper I, the system in phase II is rapidly evolving, and the loss cone is almost full for both BHs in equal mass merger. This conclusion is applicable for the secondary BHs but may not valid for the primary BHs with very light companions in unequal mass mergers. For the primary BH, a very light perturber may not be heavy enough to dominate the loss cone refilling comparing with two-body relaxation. As a result, the system will correspond to an intermediate state, which is hard to be described analytically. Thus we are focusing on the mass ratio $q\geq0.1$ in this work, which the secondary SMBH can still be considered as relatively large perturber, and the assumption of full loss cone for both BHs is still acceptable.

\subsection{Phase III}
\label{Aly-PIII}

In the last stage phase III, all of the three factors in Equation~(\ref{eq:theta_d2}) could make non-negligible contribution. The stellar distribution around an SMBHB in this stage could be triaxial \citep[e.g.][]{bort17}. And both the SMBHB and the stellar distribution are evolving. \citet{chen08} has tried to investigate tidal disruptions in compact SMBHBs with surround unbound isotropic stellar distribution by numerical scattering experiment. They have estimated the cross section for stars to be scattered to the tidal radii of both SMBHs. While the loss cone refilling rate of the SMBHB is still uncertain. In brief, the TDR is very difficult to be analytically estimated. We can only give some fitting results based on numerical simulations, which may be not convincing.

\section{Discussion}
\label{dis}

\subsection{Extrapolation}
\label{dis-extr}

As mentioned in Section~\ref{mtd}, the results of direct \nbody simulations can not be simply applied to real galaxies without extrapolation, because the limited particle resolution, which actually limited by the computer technology today, may bring some artificial effects (detailed discussions can be found in Paper I). In order to extrapolate the simulation results to reality, we need to find out how the simulation results depend on mass ratio, particle number and tidal radius. Related simulations have been done with model N500 and N250, and RT5-5 to RT1-3, respectively. Combining with simulation results for different mass ratios, it finally makes the chance to extrapolate our results to more realistic galaxies.

\subsubsection{Phase I}
\label{dis-PI}

As demonstrated by Equation~(\ref{eq:lnTDRsI}), the TDR of the primary SMBH in phase I should be independent on mass ratio. However, our result in the top left panel of Fig.~\ref{fig:qdep} shows a weak $q$ dependence. A smaller $q$ corresponds to smaller rate. That may because the situation in phase I is not exactly the same as an SMBH in an isolated galaxy. Actually, as shown in Figs.~\ref{fig:TD-std} and \ref{fig:TD-G1G2}, every time when two SMBHs approach to their pericenters in phase I, there is a small burst of the tidal disruption attendant. The loss cone refilling at those short stages should be contributed both from two-body relaxation and the companion perturbation. A model with smaller $q$ corresponds to a smaller perturber, which corresponds to smaller rate.

Taking into account contributions from all the effects discussed above, we can simply assume that the rate here is the superposition of a constant value and Equation~(\ref{eq:TDRsPIIM1}). That gives fitting results $\dot{M_1}\sim 1.8\times 10^{-5}\times q^{0.23}+1.0\times 10^{-5}$, which is denoted by red dashed line in the top left panel of Fig.~\ref{fig:qdep}. Nevertheless, since the periods around the pericenter usually are short here, this effect can only slightly change the result. In order to simplify the problem, we adopt the averaged rate $\dot{M_1} \sim 2.4\times 10^{-5}$ of all the seven models, which is indicated by the red solid line.

On the other hand, the rate of the secondary SMBH has a significant $q$ dependence, which can be well fitted by Equation~(\ref{eq:lnTDRs_2nd}) with $\dot{M_2} \sim 2.65\times 10^{-5}\times q^{0.49}(12.9+\ln q)^{0.0047}$. And there is also a good fitting result by Equation~(\ref{eq:MdotR}) in the top right panel, which gives $\dot{M_2}/\dot{M_1}\sim q^{0.40}(1+\ln q / 12.9)^{0.035}$.

\begin{figure}
\begin{center}
\includegraphics[width=0.8\textwidth,angle=0.]{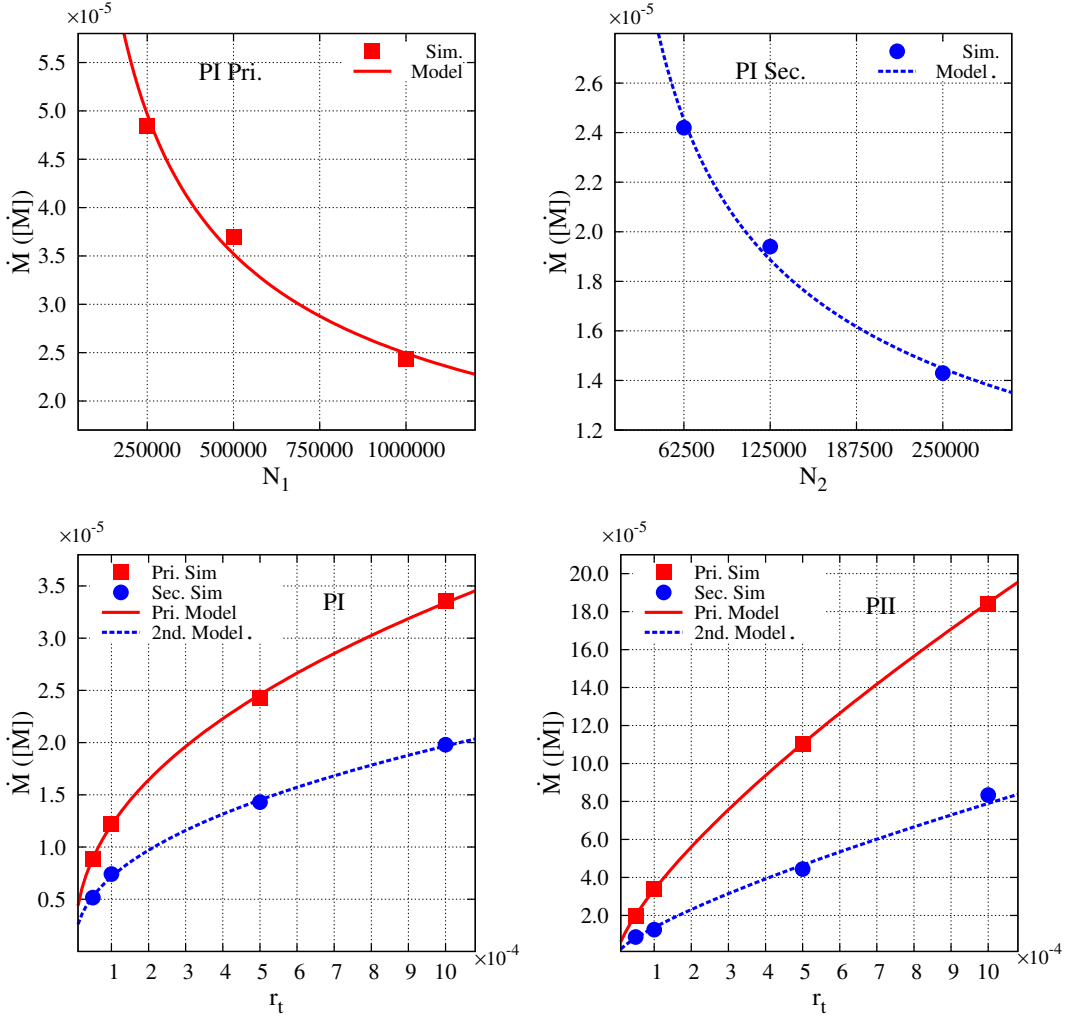}
\caption {Phase averaged tidal disruption accretion rates for different particle numbers and tidal radii. Top two panels are $N$ dependence of the primary and the secondary SMBHs in phase I. Bottom two panels are $\rt$ dependence of two SMBHs in phase I and II. The legend is the same as Fig.~\ref{fig:qdep}. (A color version is available in the online journal.)
\label{fig:Nrtdep}}
\end{center}
\end{figure}

In addition to mass ratio, according to Equations~(\ref{eq:lnTDRsI}) and (\ref{eq:lnTDRs_2nd}), if we fix other parameters, the tidal disruption accretion rate in phase I should be proportional to the power law of $N/\ln{\Lambda}$. That has been confirmed by the top panels of Fig.~\ref{fig:Nrtdep}. The top left and top right panels are, respectively, the rates for different particle numbers of the primary and the secondary SMBHs. The filled squares and circles represent our simulation results of the primary and the secondary SMBH, respectively. The solid/dashed lines are the fitting results based on Equations~(\ref{eq:lnTDRsI}) and (\ref{eq:lnTDRs_2nd}). For the primary SMBH, the fitting result gives $\dot{M_1}\sim 0.011\times(N_1/\ln{\Lambda})^{-0.54}$ in phase I, and the secondary SMBH has $\dot{M_2}\sim 9.4\times 10^{-4}\times(N_2/\ln{\Lambda'})^{-0.42}$, where $\ln{\Lambda'}=\ln(0.4N_2)$ is the Coulomb logarithm for the secondary galaxy.

The bottom left panel of Fig.~\ref{fig:Nrtdep} represents results of model RT5-5-RT1-3 and Q4 in phase I, which shows how the averaged accretion rate depends on tidal radius. It is clear that the rates in phase I of both SMBHs have significant power law dependence on $\rt$. According to Equations~(\ref{eq:lnTDRsI}) and (\ref{eq:lnTDRs_2nd}), there are good fitting results with $\dot{M_1}\sim 7.0\times 10^{-4}\times\rt^{0.44}$ and $\dot{M_2}\sim 4.1\times 10^{-4}\times\rt^{0.44}$. And the same fitted power law index for both BHs is consistent with the prediction of Equations~(\ref{eq:lnTDRsI}) and (\ref{eq:lnTDRs_2nd}).

The results above indicate that our analysis in Section~\ref{Aly-PI} for phase I can be confirmed by direct \nbody simulations. For orders of magnitude estimation, combining with simulation results, it is possible to derive the averaged tidal disruption accretion rate in phase I with different mass ratios, particle numbers and tidal radii. According to Equation~(\ref{eq:lnTDRsI}) and relevant simulation results, the rate of the primary SMBH in phase I can be estimated by a two-dimensional fitting
\beq
\ln\dot{M_1}\sim -0.97 - 0.56\ln\frac{N_1}{\ln{\Lambda}} + 0.44\ln\rt ,
\label{eq:PIM12DF}
\eeq
and the rate of the secondary SMBH in phase I, according to Equation~(\ref{eq:lnTDRs_2nd}), can be estimated by
\beq
\ln\dot{M_2}\sim -1.93 - 0.48\ln{\left(\frac{N_1}{\ln{\Lambda} + \ln{q}}\right)} + 0.51\ln{q} + 0.41\ln\rt ,
\label{eq:PIIM2MDF}
\eeq
which are consistent with previous fitting results.

Combining with Equation~(\ref{eq:scalingMdot}), the tidal disruption accretion rate in phase I for real galaxies can be derived as
\beq
\dot{M_1} \sim 1.5\times 10^{-2}\times\left(\frac{M}{10^{9}\msun}\right)^{3/2}\left(\frac{r_{1/2}}{1\kpc}\right)^{-1.94}\left(\frac{N_1}{\ln\Lambda}\right)^{-0.56}\left(\frac{\rt}{10^{-6}\pc}\right)^{0.44} \msun/\yr .
\label{eq:PIM1RLT}
\eeq
\begin{eqnarray}
\dot{M_2} &\sim & 1.1 \times 10^{-2} \times q^{0.51} \times\left(\frac{N_1}{\ln\Lambda + \ln q}\right)^{-0.48} \nonumber \\
          &\times& \left(\frac{M}{10^{9}\msun}\right)^{3/2}\left(\frac{r_{1/2}}{1\kpc}\right)^{-1.91}\left(\frac{\rt}{10^{-6}\pc}\right)^{0.41} \msun/\yr .
\label{eq:PIM2RLT}
\end{eqnarray}

The results above, especially the index of $\rt$ for the primary SMBH, however, is not the same as the Equation~(26) in Paper I. As discussed at the beginning of this subsection, the TDR for each SMBH in phase I is not exactly equivalent to the rate in isolate galaxy, because every close encounter of two nuclei could bring a small boost of tidal disruption. That means this effect weakly depends on the mass ratio. Consequently, the above results for unequal mass mergers with different mass ratios considered are not surprisingly different from Paper I based on equal mass models.

Once physical parameters of the system has been decided, combining with Equation~(\ref{eq:rt}), we can make the extrapolation. In our fiducial model, the extrapolated disruption rates of the primary and the secondary SMBH are, $\sim 7.0\times10^{-6}\msun/\yr$ and $\sim 1.1\times10^{-5}\msun/\yr$ for $q=0.25$, and $\sim 7.0\times10^{-6}\msun/\yr$ and $\sim 6.5\times10^{-6}\msun/\yr$ for $q=0.1$ respectively. It seems that the rate difference between the primary and the secondary SMBH are not significant, which is actually not unexpected. In our models the secondary galaxy always corresponds to smaller particle numbers due to the fixed configuration $N_2/N_1 = q$. While as illustrated by Fig.~\ref{fig:Nrtdep}, the rate is a monotone decreasing function of particle number. As a result, though the less massive secondary BH should correspond to significant lower rate comparing with the primary BH, the gap between two final rates after extrapolations will be narrowed because of the less star number of the secondary galaxy/nucleus.

\citet{liu13} have analytically estimated the TDR for different minor mergers. We have compared our result of the fiducial model in phase I with their model. Since \citet{liu13} did not present their results for the mass ratio $q = 0.25$, we follow their model to have a new integration. With similar parameters adopted in our fiducial model, their model gives the TDE rates for the primary SMBH and the secondary SMBH are $1.9\times10^{-5}/\yr$ and $3.4\times10^{-5}/\yr$, respectively. Considering that our extrapolation is just the orders of magnitude estimation, there is a very good consistency when two galaxies are far away from each other. However, their analytical estimation tends to give higher rate when two SMBHs are getting close in phase I. We believe here are two reasons. First of all, \citet{liu13} have assumed spiral-in orbits, which corresponds to long term gradual evolution. While in our models two SMBHs with their star clusters are in parabolic orbits with relatively rapid orbital evolution, corresponding to much shorter interaction periods. Secondly, we assumed that the loss cone refilling are dominated by two-body relaxation in phase I and by perturbation in phase II. However, due to limited particle resolution, the mass ratio between SMBHs and stars in our simulation models are significantly smaller than they should have in real galaxies. Thus the two-body relaxation are artificially enhanced in our simulation. Consequently, the transition from phase I to phase II in our simulation tend to be delayed comparing with reality. Then two SMBHs need to get more closer to enter the perturbation dominate phase in our simulations. Though above extrapolations could avoid the influence of limited particle number and tidal radius to averaged accretion rates, it is still insufficient to derive accurate transition boundary between phase I and II. Nevertheless, on orders of magnitude, our extrapolations on typical averaged accretion rates in phase I above and phase II bellow should be still valid.

\subsubsection{Phase II}
\label{dis-PII}

In phase II, the rate of the primary SMBH should weakly depend on $q$ according to Equation~(\ref{eq:TDRsPIIM1}), while the rate of the secondary SMBH should be proportional to $q^\alpha$. Our results in the bottom left panel of Fig.~\ref{fig:qdep} confirm that prediction. The dependence of the primary SMBH on $q$ is very weak, which gives a fitting result $\dot{M_1}\sim 1.1\times 10^{-4}\times q^{0.009}$, as indicated by red solid line. And the relation between the rate of the secondary SMBH and the mass ratio can be fitted by $\dot{M_2}\sim 1.04\times 10^{-4}\times q^{0.57}$, which is indicated by blue dashed line.

As shown in Table~\ref{tab:res-all}, since the loss cone of both SMBHs are relatively full, the rates of two SMBHs do not significantly depend on particle numbers in phase II. While as represented in the bottom right panel of Fig.~\ref{fig:Nrtdep}, the results of the primary and the secondary SMBHs can also be fitted quite well by Equations~(\ref{eq:TDRsPIIM1}) and (\ref{eq:TDRsPIIM2}), respectively. For the primary SMBH it is $\dot{M_1}\sim 3.1\times 10^{-2}\times\rt^{0.74}$, and the secondary SMBH has $\dot{M_2}\sim 1.6\times 10^{-2}\times\rt^{0.76}$.

Analog to phase I, we can make the similar extrapolations for phase II. Since the rates of both SMBHs are not $N$ dependent, they can be estimated through Equations~(\ref{eq:TDRsPIIM1}), (\ref{eq:TDRsPIIM2}) and the fitting results above as
\beq
\ln\dot{M_1}\sim -3.61 + 0.007\ln{q} + 0.73\ln\rt ,
\label{eq:PIIM12DF}
\eeq
\beq
\ln\dot{M_2}\sim -3.30 + 0.56\ln{q} + 0.77\ln{\rt}.
\label{eq:PIIM22DF}
\eeq
And the corresponding estimation for real galaxies are
\beq
\dot{M_1} \sim 3.5\times 10^{-6}q^{0.007}\times\left(\frac{M}{10^{9}\msun}\right)^{3/2}\left(\frac{r_{1/2}}{1\kpc}\right)^{-2.23}\left(\frac{\rt}{10^{-6}\pc}\right)^{0.73} \msun/\yr .
\label{eq:PIIM1RLT}
\eeq
\beq
\dot{M_2} \sim 2.1\times 10^{-6}q^{0.56}\times\left(\frac{M}{10^{9}\msun}\right)^{3/2}\left(\frac{r_{1/2}}{1\kpc}\right)^{-2.27}\left(\frac{\rt}{10^{-6}\pc}\right)^{0.77} \msun/\yr .
\label{eq:PIIM2RLT}
\eeq

In our fiducial model, the averaged disruption rates in phase II for the primary and the secondary SMBH are, $\sim 1.3\times10^{-4}\msun/\yr$ and $\sim 3.9\times10^{-5}\msun/\yr$ for $q=0.25$, and $\sim 1.3\times10^{-4}\msun/\yr$ and $\sim 2.3\times10^{-5}\msun/\yr$ for $q=0.1$ respectively. The corresponding peak rates at $q=0.25$ are $\sim 3.6\times10^{-4}\msun/\yr$ and $\sim 8.3\times10^{-5}\msun/\yr$, respectively, for the primary and the secondary SMBH. And the peak rates at $q=0.1$ are $\sim 3.6\times10^{-4}\msun/\yr$ and $\sim 6.1\times10^{-5}\msun/\yr$, respectively, for the primary and the secondary SMBH. In general, the averaged TDR in phase II can be enhanced several times to an order of magnitude relative to phase I. And the peak rate in phase II can be boosted dozens of times.

In phase II, the separation of two SMBHs shrinks from several times of the primary SMBH's influence radius to a few percent within very short time. The estimations made by \citet{liu13} is not valid here, because they integrate the tidal disruption evolution to the separation around twice influence radius, which only corresponds to the beginning of phase II here. \citet{chen09} and \citet{chen11} have investigated the TDR evolution for close SMBHBs. They found that the tidal disruption contributed by both Kozai-Lidov effect \citep{koz62,lid62} and chaotic loss cone repopulation. For the binary with high eccentricity, which is the situation we have here, the latter is dominated. However, since they focused on systems with very small mass ratio, which neglected the stellar distribution distortion induced by the two SMBHs, comparison between their results and ours is not straightforward.

\subsubsection{Phase III}
\label{dis-PIII}

We have tried to numerically fit the q dependence results and find that the rate of the primary SMBH can be fitted by a power law $\dot{M_1}\sim 3.0\times 10^{-5}\times q^{0.47}$, and the fit to the rate of the secondary SMBH prefer a linear relation, which gives $\dot{M_2}\sim 2.7\times 10^{-5}\times q+1.8\times 10^{-6}$. Probably our assumption of a scattering process for loss cone refilling of the secondary SMBH in phase II is more or less still valid for phase III.

Similar to phase II, we can not find significant $N$ dependence for both SMBHs. However, due to our limited knowledge about how the loss cone refilling process depends on simulation particle resolution in phase III, it should be aware that the particle dependence results here may not be an universal answer. On the other hand, as presented by Fig.~\ref{fig:rtPIII} in Appendix~\ref{sec:app}, though the dependence of TDRs on $\rt$ is not analytically clear in phase III, we still found that the tidal radius dependence of the primary SMBH can be roughly fitted by a power law with $\dot{M_1}\sim 3.6\times 10^{-3}\times\rt^{0.73}$, while the secondary SMBH can be fitted by a linear relation with $\dot{M_2}\sim 1.3\times 10^{-2} \times \rt + 9.2\times 10^{-7}$.

As mentioned before, there is no analytical results to guide us to finish the extrapolation. We just assume that the rate dependence on $N$ and $\rt$ is similar as phase II. It can be fitted as

\beq
\ln\dot{M_1}\sim -5.1 + 0.48\ln{q} + 0.70\ln\rt ,
\label{eq:PIIIM12DF}
\eeq
\beq
\ln\dot{M_2}\sim -4.42 + 0.63\ln{q} + 0.84\ln{\rt} ,
\label{eq:PIIIM22DF}
\eeq
Which gives
\beq
\dot{M_1} \sim 6.3\times 10^{-6}q^{0.48}\times\left(\frac{M}{10^{9}\msun}\right)^{3/2}\left(\frac{r_{1/2}}{1\kpc}\right)^{-2.20}\left(\frac{\rt}{10^{-6}\pc}\right)^{0.70} \msun/\yr .
\label{eq:PIIIM1RLT}
\eeq
\beq
\dot{M_2} \sim 5.4\times 10^{-7}q^{0.63}\times\left(\frac{M}{10^{9}\msun}\right)^{3/2}\left(\frac{r_{1/2}}{1\kpc}\right)^{-2.34}\left(\frac{\rt}{10^{-6}\pc}\right)^{0.84} \msun/\yr .
\label{eq:PIIIM2RLT}
\eeq

In the fiducial model with $q=0.25$, the averaged rates in phase III are $\sim 1.1\times10^{-4}\msun/\yr$ and $\sim 1.0\times10^{-5}\msun/\yr$ for the primary and the secondary SMBH, respectively. However, it should be aware that, though our extrapolations purely based on numerical simulation results are possible here, the rate estimations on phase III are still not robust enough without the guideline of theoretical analysis.

\subsection{Observation Implications}
\label{dis-obs}

As mentioned in Paper I, the boosted TDR during phase II contributes more than a quarter of disruption events in equal mass mergers. However, that may not be the case for minor mergers. According to our discussion in Section~\ref{dis-extr}, the averaged TDRs in phase II usually are an order of magnitude higher than phase I, which may indicate that the contribution in phase II should be significant. But as demonstrated in Fig.~\ref{fig:TD-all} and Table~\ref{tab:res-time}, those smaller mass ratio models usually correspond to significantly extended periods of phase I, while periods of phase II for all the models are roughly the same. If we assume that the evolution of two SMBHs before phase II are dominated by dynamical friction, and the corresponding TDR is approximately equal to the averaged TDR in phase I, then the total TDE number will be proportional to $\Gamma\times t_{\rm df}$, where $\Gamma$ is the tidal disruption event rate, and $t_{\rm df}\propto q^{-1}$ is dynamical friction timescale \citep{liu13}. For a minor merger with $q \lesssim 0.25$, comparing with equal mass merger, it may take several times longer duration to evolve to phase II. As a result, the contribution of phase II in minor mergers could be submerged by the earlier and later stages, with lower rates but longer periods. In other words, in major mergers a detected TDE still has relatively high possibility to be occurred in phase II, while the possibility of a TDE detected in a minor merger which contributed by phase II should be small. However, the minor mergers are more common. According to \citet{lotz11}, the minor merger rate ($0.1\lesssim q < 0.25$) in local universe is $\sim 3$ times of major mergers. Actually, based on detailed predictions about the SMBHB induced TDE detection rate made by \citet{thor18}, the Large Synoptic Survey Telescope (LSST)\footnote{https://www.lsst.org/lsst} and the extended ROentgen Survey with an Imaging Telescope Array (eROSITA)\footnote{https://www.mpe.mpg.de/eROSITA} may detect tens of TDEs induced by SMBHB every year. That means the we may still have chance to detect a few TDEs corresponding to phase II stage of minor mergers every year. Another planed X-ray space transients survey, the Einstein Probe (EP)\footnote{http://ep.bao.ac.cn/} can also detect several tens to hundreds TDEs per year \citep{yuan15}, which corresponds to a few detections relevant to phase II of minor mergers within the lifetime of the satellite. It should be aware that, according to the discussion in Section~\ref{dis-extr}, our results tend to underestimate the duration of phase II. Thus above estimation of TDEs in phase II can only give a lower limit.

An interesting result is that the TDR of the secondary SMBH in phase I can be close to the primary SMBH for most of minor merger models. As mentioned in Section~\ref{dis-extr}, smaller galaxies with less stars correspond to shorter two-body relaxation timescales, which results in relatively higher TDRs in phase I. In other words, the contribution of both SMBHs in a minor merger is similar in the early stage. In phase II, though the contribution is dominated by the primary SMBH, there is still significant fraction of TDEs contributed by the secondary SMBH. Furthermore, there is another special but not rare condition, that the primary SMBH has mass exceed $\sim 10^8 \msun$. In that case the tidal radius of the SMBH for solar type stars is likely smaller than its Schwarzschild radius, which means the tidal disruptions from the primary SMBH may do not have emission at all. Then the TDEs of the secondary SMBH will dominate the contribution.

Besides, according to Fig.~\ref{fig:TD-G1G2}, most of tidal disruptions of the secondary SMBH are contributed by stars from the other galaxy in phase II and III. That indicates many of those disrupted stars could have parabolic or even hyperbolic orbits before tidal disruption. Most of observations on tidal disruption right now, however, have assumed that disrupted stars are in parabolic orbits. In reality, the light curve of a TDE from eccentric or hyperbolic star can be significantly different from the "classic" parabolic orbits \citep{hay13,hay18}. As extreme cases, all the debris from disrupted stars with strongly bound eccentric orbits will fall back to the SMBH, leading to a significant deviation from the "canonical" $t^{-5/3}$ fall back rate. And there will be no debris fall back from those disrupted stars with extreme hyperbolic orbits, which corresponds to invisible tidal disruption events. Recently, \citet{hay18} found that almost all of disrupted stars are in parabolic orbits for a single SMBH in a spherical isolate galaxy. While our results here tends to indicate that the disrupted stars with hyperbolic orbits may occupy a large fraction in merging galaxies. Actually we also find significant fraction of stars with eccentric orbits in our simulation results. Detailed discussion about this issue is beyond the scope of this paper. We will have carefully investigations in the next paper.

\section{Summary}
\label{sum}

We have done direct \nbody simulations of unequal mass merging galaxies with supermassive black holes (SMBH) in their centers, with a detailed study of the tidal disruption rates (TDR) of stars near the SMBH; the evolution is divided into three phases, first "phase I", where both SMBH and their nuclear star clusters are largely independent of each other, "phase II" where a bound supermassive black hole binary (SMBHB) forms, and there is a strong interaction of both nuclear star clusters, rapid and presumably chaotic evolution with a significantly enhanced TDR, and a "phase III" during which the system has settled. We find that the TDR during the formation of a bound SMBHBs, can be magnified several times to dozens of times, and there are power law relations between TDR and mass ratio of two SMBHs/galaxies. In addition, we also find that in a minor merger, after two SMBHs become bound, TDEs of both SMBHs are mainly contributed by stars from the more massive galaxy.

Based on the method and results in Paper I, here we have extended our research on tidal disruption in galaxy mergers to more popular and realistic unequal mass mergers. Though the dynamical evolution timescales are longer, their TDR evolutions are quite similar to equal mass mergers. There will be significantly enhanced disruption rates in phase II, with maximum rates boosted to dozens of times relative to normal galaxies. The rate of the primary SMBH weakly or barely depends on the mass ratio before and during the formation of the SMBHB. While the rate of the secondary SMBH has significant mass ratio dependence, which gives $\dot{M_2} \sim 2.65\times 10^{-5}\times q^{0.49}(12.9+\ln q)^{0.0047}$ in phase I, and $\dot{M_2}\sim 1.04\times 10^{-4}\times q^{0.57}$ in phase II. The rates for both SMBHs obviously depend on mass ratio after a compact SMBHB has been formed. More interestingly, during the late stage of phase II and the early stage of phase III, most of stars disrupted by the secondary SMBH actually originate from the host galaxy/nucleus of the other SMBH. That means there should be some unbound disrupted stars in hyperbolic orbits, which has significantly different light curves compare with classical tidal disruptions with parabolic orbits assumed.

As direct \nbody simulations, the results inevitably limited by the particle resolution. Due to the high accuracy requirement of the integration, none of the high performance cluster in the world can afford a direct \nbody simulation with particle number close to the number of stars in a normal galaxy. An alternative solution is to simulate a galaxy with much less particles and then extrapolate results to the situation in reality, which needs to be very careful because many physical processes, for instance two-body relaxation, significantly depend on the particle number. In Paper I, by dividing the whole evolution into three different phases, we analytically assumed some dominate process in each phase, and derived corresponding $N$ and $\rt$ dependence. With these guidelines, combined with simulation results in different parameters, we made convincing extrapolations and derive the TDRs for real equal mass mergers. As demonstrated in Sections~\ref{dis-extr}, this strategy is also valid for minor mergers. We analytically estimate the $N$, $\rt$ and $q$ dependence of TDR in phase I and II, and found that they are consistent with our simulation results. Therefore, combined the theoretical analyses and direct \nbody simulation results, we can estimate the TDRs of minor mergers for the first time. The only deficiency is that the rate in phase III is still not clear, because it is contributed by many processes and none of them is dominated.

\acknowledgments

We are grateful to Alberto Sesana and Shiyan Zhong for helpful discussions. This work is supported by the National Natural Science Foundation of China (NSFC11303039, NSFC11673032), the Key International Partnership Program of the Chinese Academy of Sciences (CAS) (No.114A11KYSB20170015), and the Strategic Priority Research Program (Pilot B) Multiwavelength gravitational wave universe of CAS (No.XDB23040100). We (SL, PB, RS) acknowledge support by CAS through the Silk Road Project at National Astronomical Observatories (NAOC) with the ¡°Qianren¡± special foreign experts program of China, and the support by Key Laboratory of Computational Astrophysics. The computations have been done on the Laohu supercomputer at the Center of Information and Computing at NAOC, CAS, funded by Ministry of Finance of People's Republic of China under the grant $ZDYZ2008-2$. PB acknowledges the special support by the CAS President's International Fellowship for Visiting Scientists (PIFI) program during his stay in NAOC, CAS. FKL and XC acknowledge the support by the National Natural Science Foundation of China (NSFC11473003, NSFC11873022).

\appendix

\section{Analytical model for equal mass mergers}
\label{sec:app}

As demonstrated in Paper I, it is possible to analytically estimate the TDR evolution for equal mass mergers. A single star which is approaching a single BH will be tidally disrupted within one orbital period if its specific angular momentum is less than $J_{\rm {lc}}\simeq(2G\Mbh\rt)^{1/2}$. All of such kind of possible orbits in phase space are inside a cone region, which is the so-called "loss cone". The velocity vectors corresponding to loss cone has an opening angle\citep[][and references therein]{lig77, merr13}
\beq
\thetalc = \frac{J_{\rm lc}}{J_{\rm c}},
\label{eq:theta_lc}
\eeq
where $J_{\rm c}$ is the specific circular angular momentum with same energy of the star.

In a spherical isotropic system with equal stellar mass, $\thetalc$ can be estimated inside the influence radius $\rinf$ of the BH,
\beq
\theta^{2}_{\rm{lc}} \simeq \frac{2}{3}\frac{r_{\rm t}}{r},
\label{eq:theta_lc2a}
\eeq
\citep{fra76,bau04}.
In a more realistic system with lots of stars, the angular momentum of every star changes with time because of gravitational interactions between them. If the effective deflection angle is small enough with $\thetaD(r)\ll\thetalc(r)$, which means the orbital averaged deflection angle of velocity is very small, all the stars inside loss cone will be disrupted by BH within one orbital period \citep{fra76,shap76,lig77}. As a result, the loss cone will be empty because of the inefficient replenishment of stars through diffusion. Conversely, if $\thetaD(r)\gg\thetalc(r)$, the replenishment of stars is so efficient that the loss cone can keep full. Both theoretical and numerical estimations find that there will be a critical radius $\rcrit$ corresponding to $\thetaD(r_{crit})\sim\thetalc(r_{crit})$ \citep{fra76,lig77,amar04,zho14}.

In a spherical system, the TDE rate within a shell can be estimated on the order of magnitude by
\beq
d\Gamma \backsimeq \frac{4\pi r^2 \rm dr\rho(r)}{m_*}\frac{\theta^2(r)}{t_{\rm d}(r)}
\label{eq:TDNRT}
\eeq
\citep{fra76,syer99,liu13}, where $t_{\rm d}(r)$ is dynamical timescale of the star, $\rho$ is stellar mass density, and
\beq
\theta^2=\min(\thetalc^2, \thetaD^2/\ln\thetalc^{-1})
\label{eq:thetasq}
\eeq
is a dimensionless coefficient for stars depleting into the loss cone\citep{syer99}. In order to precisely calculate the total disruption rate, one should integrate this formula. However, for order of magnitude estimation, we can assume that TDEs are mainly contributed by stars around the critical radius, which gives
\beq
\Gamma \sim \left. \frac{r^3\theta^{2}_{\rm{D}}\rho(r)}{t_{\rm d}\mstar}\right|_{r=r_{\rm{crit}}}
\label{eq:TDrate}
\eeq
\citep{fra76,bau04}, where for simplicity all stars are assumed to have equal mass ($\mstar=\msun$).

In order to analytically estimate the evolution of TDR in merging galaxies, \citet{liu13} assumed that the effective deflection angle is mainly contributed by stellar two-body relaxation, massive perturbation from another SMBH with surrounding stars, and triaxial stellar distribution
\beq
\thetaD^2 = \theta_{\rm 2}^2 + \theta_{\rm p}^2 + \theta_{\rm c}^2 ,
\label{eq:theta_d2}
\eeq
where $\theta_{\rm 2}$, $\theta_{\rm p}$ and $\theta_{\rm c}$ are, respectively, the contribution from two-body relaxation, massive perturber, and triaxial gravitational potential. According to the separation of two SMBHs, and results based on direct \nbody simulation for equal mass mergers in Paper I, the evolution of merging systems can be divided to three phases. In phase I, two merging stellar systems still keep relatively large distance, which means the perturbation from one system to the other is not significant. For order of magnitude estimation, we can assume most of TDEs in this stage are contributed by two-body relaxation, which corresponds to $\theta_{\rm D} \sim \theta_{\rm 2}$. In phase II, two systems are so close that two SMBHs will slip into the influence radii of each other and finally form a binary system. The period of this phase is relatively short and the system contains lots of stars with chaotic orbits. For this reason, the perturbation dominates the contribution to TDEs. After an SMBHB formed, it will be continuously hardening in phase III. The estimation of TDE in this stage is very complicated, because all three factors listed above may have non-negligible contributions.

\subsection{Phase I}
\label{sec:appPI}

In phase I, as mentioned in Paper I, due to the relatively large separation, the tidal disruption rate of each galaxy is more or less similar to isolate galaxy, which means two-body relaxation dominates the loss cone refilling. For this reason, with assumption $\theta_{\rm D} \sim \theta_{\rm 2}$, the mass accretion rate of each SMBH due to tidal disruption can be described by following equations, which are the same as Equations~$(23)$ and $(24)$ in Paper I,
\beq
\ln\dot{M_1}\approx \ln A - \frac{2\gamma-1}{8-2\gamma}\ln\left(\frac{N_1}{\ln\Lambda}\right) + \frac{9-4\gamma}{8-2\gamma}\ln\rt,
\label{eq:lnTDRs}
\eeq
and
\beq
A = 2.94\times 0.23^{\frac{9-4\gamma}{8-2\gamma}}G^{\frac{1}{2}}\left(\rho_0r_0^\gamma\right)^{\frac{7}{8-2\gamma}}\Mbhh^{\frac{6-5\gamma}{8-2\gamma}}.
\eeq
Where$\ln \Lambda \thickapprox \ln (0.4N_1)$ is the Coulomb logarithm for the primary galaxy \citep{spit87}. Here we assume that the mass of the SMBH will not be significantly changed by mass accretion due to tidal disruption, and $\rho(r) = \rho_0 (r/r_0)^{-\gamma}$.

\subsection{Phase II}
\label{sec:appPII}

In phase II, which roughly corresponds to the stage where the companion BH enters the influence radius of the primary BH, the perturbation of the companion BH dominates the replenishing process of the loss cone. Thus the estimation in Paper I is still valid. According to Equation~(31) in Paper I, and neglecting the q dependence for $t_{\rm \omega}$ in that equation, the tidal disruption accretion rate of the primary BH is
\beq
\dot{M}_1 \propto M_{\rm p}^{\frac{4-2\gamma}{11}}\Mbh^{\frac{7-3\gamma}{11}}\rt^{\frac{12-2\gamma}{11}},
\label{eq:TDRsPII}
\eeq
where $M_{\rm p}$ is the total mass of the perturber.

\subsection{Phase III}
\label{sec:appPIII}

\begin{figure}
\begin{center}
\includegraphics[width=0.8\textwidth,angle=0.]{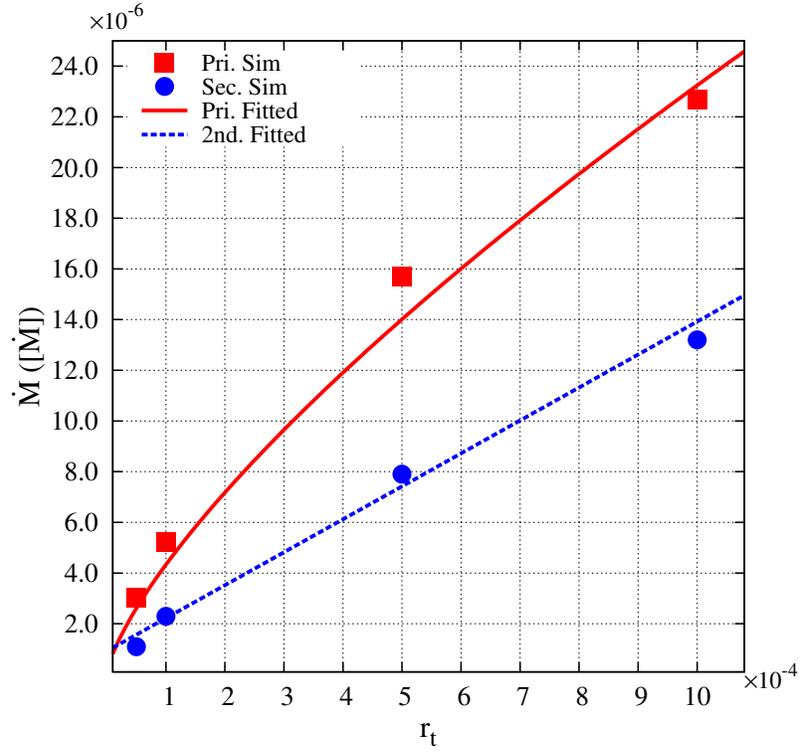}
\caption {Phase averaged tidal disruption accretion rates for different tidal radii in phase III. The legend is the same as Fig.~\ref{fig:qdep}. (A color version is available in the online journal.)
\label{fig:rtPIII}}
\end{center}
\end{figure}

Since all of the three factors may have significant contributions, the estimation of TDR in phase III is quite uncertain. Therefore we will not give any analytical estimations. However, it is possible to get some dependencies by purely numerical fittings. As shown in Figs.~\ref{fig:rtPIII}, there are significant $\rt$ dependencies for both SMBHs. More details can be found in Section~\ref{dis-PIII}.

\end{document}